\documentclass[prd,showpacs,preprintnumbers,amsmath,amssymb,superscriptaddress,floatfix,nofootinbib]{revtex4}
\usepackage{graphicx}
\usepackage{subfigure}
\usepackage{amsmath}
\usepackage{amsfonts}
\usepackage{amssymb}
\usepackage{color}
\usepackage{multirow}
\usepackage[colorlinks, citecolor=blue,anchorcolor=red,menucolor=red,linkcolor=red,filecolor=red,runcolor=red,urlcolor=blue,frenchlinks=red]{hyperref}

\begin{document}

\title{Enhancement near the $\bar{p}\Lambda$ threshold in the $\chi_{c0}\to \bar{p}K^+\Lambda$ reaction}

\author{Guan-Ying Wang}
\affiliation{School of Physics and Microelectronics, Zhengzhou University, Zhengzhou, Henan 450001, China}
\affiliation{School of Physics and Electronics, Henan University, Kaifeng 475004, China}
\affiliation{International Joint Research Laboratory of New Energy Materials and Devices of Henan Province, Henan University, Kaifeng 475004, China}

\author{Man-Yu Duan}
\affiliation{School of Physics and Microelectronics, Zhengzhou University, Zhengzhou, Henan 450001, China}

\author{En Wang}
\email{wangen@zzu.edu.cn}
\affiliation{School of Physics and Microelectronics, Zhengzhou University, Zhengzhou, Henan 450001, China}

\author{De-Min Li}
\email{lidm@zzu.edu.cn}
\affiliation{School of Physics and Microelectronics, Zhengzhou University, Zhengzhou, Henan 450001, China}

\begin{abstract}
We have analyzed the reaction $\chi_{c0}\to \bar{p} K^+\Lambda$ reported by the BESIII Collaboration, taking into account the contributions from the intermediate $K(1830)$, $N(2300)$, and $\Lambda(1520)$ resonances. Our results are in good agreement with the BESIII measurements, and it is found that the anomalous enhancement near the $\bar{p}\Lambda$ threshold is mainly due to the contribution of the $K(1830)$ resonance. We also show that the interference of the high-mass $N^*$ and $\Lambda^*$ can not produce the anomalous enhancement near the $\bar{p}\Lambda$ threshold.
\end{abstract}


\maketitle

\section{Introduction}
\label{sec:introduction}
The hadronic decays of the charmonium states could be used to understand the mechanisms of the charmonium decays, and provide a good place to search for the light baryons and mesons, since the charmonium states are the SU(3) singlets, and the final states could provide an isospin filter~\cite{Klempt:2007cp,Zou:2000wg,Ablikim:2012ih, Ablikim:2012jg,Ablikim:2012hi,Ablikim:2011uf,Ablikim:2014dnh}. For instance, we have studied the reactions of $\chi_{c0}\to \bar{\Sigma}\Sigma \pi$ and $\chi_{c0}\to \bar{\Lambda}\Sigma \pi$, which could be used to search for the baryon state $\Sigma(1430)$ with $J^P=1/2^-$ and to understand the two-pole structure of the $\Lambda(1405)$ resonance~\cite{Wang:2015qta,liu:2017efp}.

In 2013, an anomalous enhancement near the $\bar{p}\Lambda$ threshold was observed by the BESIII Collaboration in the $\chi_{c0}\to \bar{p}K^+\Lambda$ process~\cite{Ablikim:2012ff}. Assuming the relative angular momentum $L=0$ between $\bar{p}$ and $\Lambda$, the BESIII Collaboration made a fit to the data of the $\bar{p}\Lambda$ mass distribution, and give a state with $M=2053\pm 13$~MeV and $\Gamma=292\pm 14$~MeV~\cite{Ablikim:2012ff}.
On the other hand, the similar anomalous enhancements near the $\bar{p}\Lambda$ (or $p\bar\Lambda$) threshold were also observed in other processes, such as the $J/\psi \to p K^-\bar{\Lambda}+c.c.$, $\psi' \to p K^-\bar{\Lambda}+c.c.$~\cite{Ablikim:2004dj}, $B^0\to p \bar{\Lambda} \pi^-$~\cite{Wang:2003yi}, $B^-\to J/\psi \Lambda \bar{p}$~\cite{Xie:2005tf},  and $\psi(3680)\to \gamma\chi_{cJ}\to \gamma \bar{p}K^{*+}\Lambda+c.c.$~\cite{Ablikim:2019sve}.

 Most often an enhancement close to the threshold is an indication of the bound state or resonance below threshold~\cite{Aceti:2014kja,Wang:2019evy}. For instance, a peak observed in the $\phi\omega$ threshold in the $J/\psi\to \gamma \phi\omega$ reaction~\cite{Ablikim:2006dw} was interpreted as the manifestation of the $f_0(1710)$ resonance below the $\phi\omega$ threshold~\cite{Geng:2008gx}. In Ref.~\cite{Ablikim:2009ac} the BESIII Collaboration has seen a bump structure close to threshold in the $K^{*0}\bar{K}^{*0}$ mass distribution of the $J/\psi\to \eta K^{*0}\bar{K}^{*0}$ decay, which can be interpreted as a signal of the formation of an $h_1$ resonance~~\cite{Xie:2013ula,Geng:2008gx}.

 The nature of the anomalous enhancement near the $\bar{p}\Lambda$ (or $p\bar{\Lambda}$ ) threshold is not clear. The anomalous enhancement near the $\bar{p}\Lambda$ threshold may be interpreted as a quasibound dibaryon, or simply as an interference effect of high-mass  $N^*$ and $\Lambda^*$ states, as mentioned by Ref.~\cite{Ablikim:2012ff}.  By investigating the $p\bar{\Lambda}$ systems of $J=0,1$ within the chiral quark model and the quark delocalization color screening model, Ref.~\cite{Huang:2011zq} has shown that there is no $S$-wave bound state.
On the other hand, a preliminary study in the chiral effective field theory of Refs.~\cite{Li:2016mln,Song:2018qqm} showed that the $S$-wave $\bar{p}\Lambda$ interaction is weak and could not generate a bound state~\cite{geng}.
The enhancement near the $\bar{p}\Lambda$ threshold seems unlikely to be a quasibound dibaryon.
In addition, the partial wave analysis performed by Ref.~\cite{Ablikim:2004dj} has shown that the enhancement near the $p\bar\Lambda$ threshold in the $J/\psi\to p K^-\bar{\Lambda}$ process cannot be due to the high-mass $N^*$ and $\Lambda^*$ interference effect. One purpose of this work is to check whether the enhancement structure near the $\bar{p}\Lambda$ threshold in the $\chi_{c0}\to \bar{p}K^+\Lambda$ process can be interpreted as the  high-mass $N^*$ and $\Lambda^*$ interference effect or not. Also, some of the high-mass excited kaon states such as the $K_2(2250)$, $K_3(2320)$, and $K_4(2500)$ have been observed in the $\bar{p}\Lambda$ (or $p\bar{\Lambda}$) mode~\cite{PDG2018}, which implies that the high excited kaons could couple to the $\bar{p}\Lambda$ (or $p\bar{\Lambda}$) channel. One can naturally ask whether the anomalous enhancement near the $\bar{p}\Lambda$ threshold is due to the high-mass excited kaon states or not.
 We would like to propose that the enhancement near the $\bar{p}\Lambda$ threshold in the $\chi_{c0}\to \bar{p}K^+\Lambda$ process may be an indication of the excited kaon below the $\bar{p}\Lambda$ threshold. This is another purpose of this work.

Based on the fact that the $\bar{p}K^+$ mass distribution has a clear peak around 1520~MeV associated to the $\Lambda(1520)$ state and the $\Lambda K^+$ shows a peak structure around $2200\sim2300$~MeV associated to the $N^*$ states~\cite{Ablikim:2012ff}, we will consider the contributions from the intermediate $\Lambda(1520)$ and $N^*$ resonances in the $\chi_{c0}\to \bar{p}K^+\Lambda $ reaction. In addition, we will consider the contribution from the excited kaons in this reaction.

This paper is organized as follows. In Sec.~\ref{sec:formalism}, we will present the mechanism for the reaction of $\chi_{c0}\to \bar{p}K^+\Lambda$, and in Sec.~\ref{sec:results}, we will show our results and discussion. Finally a summary is given in Sec.~\ref{sec:summary}.

\section{Formalism}
\label{sec:formalism}
\begin{figure}[h]
\begin{center}
\includegraphics[width=0.45\textwidth]{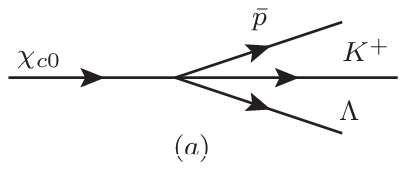}
\includegraphics[width=0.45\textwidth]{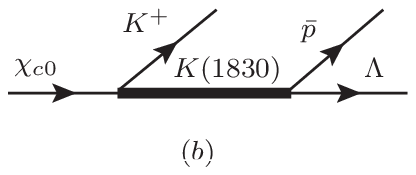}
\includegraphics[width=0.45\textwidth]{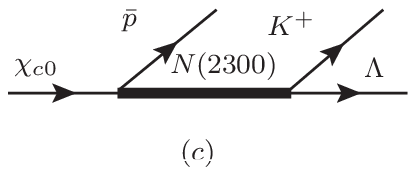}
\includegraphics[width=0.45\textwidth]{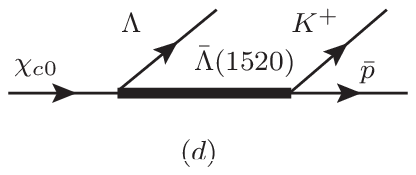}
\caption{The diagram for the reaction $\chi_{c0}\to \bar{p} K^+\Lambda$, (a) the direct diagram, (b) the contribution from the intermediate $K(1830)$ resonance, (c) the contribution from the intermediate $N(2300)$ resonance, and (d) the contribution from the intermediate $\Lambda(1520)$ resonance.}
\label{fig:feyn}
\end{center}
\end{figure}

In this section, we will present the mechanism for the reaction $\chi_{c0}\to \bar{p}K^+\Lambda $.  In addition to the direct diagram of Fig.~\ref{fig:feyn}(a), we take into account the contribution from the intermediate excited kaon (denoted as $K^*$ below), as shown in Fig.~\ref{fig:feyn}(b).
According to the PDG~\cite{PDG2018}, there are several $K^*$ states close to the $\bar{p}\Lambda$ threshold, such as the $K(1830)$, $K_2(2250)$ and $K_3(2320)$, however, only the $K(1830)$ [$I(J^P)=1/2(0^-)$] could couple to the $\bar{p}\Lambda$ in $S$-wave, and the vertex $\chi_{c0}\to K\bar{K}(1830)$ is also in $S$-wave. We thus only consider the contribution from the intermediate $K(1830)$ state in the present work because the contributions from the higher angular momentum hypotheses are expected to be strongly suppressed near threshold.

From the measurements of the $\chi_{c0}\to \bar{p}K^+\Lambda $ reaction shown in Fig.~6(a) of Ref.~\cite{Ablikim:2012ff}, one can find
a clear peak around 1520~MeV in the $\bar{p}K^+$ mass distribution, associated to the $\Lambda(1520)$ state, and a broad peak around $2200\sim 2300$~MeV in the $\Lambda K^+$ mass distribution, which corresponds to the intermediate $N^*$ states. In this work, we consider the contribution from the $N(2300)$ resonance which could couple to the $\Lambda K^+$ in $P$-wave, as shown in Fig.~\ref{fig:feyn}(c), although there are four $N^*$ states in this region [$N(2190)$ $(7/2^-)$, $N(2220)$ $(9/2^+)$, $N(2250)$ $(9/2^-)$, and $N(2300)$ $(1/2^+)$]~\cite{PDG2018}, and also the one from the $\Lambda(1520)$ as shown in Fig.~\ref{fig:feyn}(d).

The total amplitude for the $\chi_{c0}\to \bar{p}K^+\Lambda $ reaction considered in this work can be written as
\begin{equation}
\mathcal{M}^{\rm total}=\mathcal{M}^{\rm direct}+\mathcal{M}^{K(1830)}+\mathcal{M}^{N(2300)}+\mathcal{M}^{\Lambda(1520)}, \label{eq:fullamp}
\end{equation}
where the $\mathcal{M}^{\rm direct}$, $\mathcal{M}^{K(1830)}$, $\mathcal{M}^{N(2300)}$, and $\mathcal{M}^{\Lambda(1520)}$ are the amplitudes from the direct diagram, $K(1830)$, $N(2300)$, and $\Lambda(1520)$, respectively.

The amplitude of the direct diagram and the $K(1830)$ term can be expressed as
\begin{eqnarray}
\mathcal{M}&=&\mathcal{M}^{\rm direct} +\mathcal{M}^{K(1830)} \nonumber \\
&=& V_p \left[1+ \frac{\alpha M^2_N}{M_{\bar{p}\Lambda}^2-M_{K^*}^2+i M_{K^*}{\Gamma_{K^*}}} \right], \label{eq:amp_kstar}
\end{eqnarray}
where $V_p$ is an unknown normalization factor, $\alpha$ is the weight of the contribution from the intermediate $K(1830)$ state with a mass of $M_{K^*}$ and a width of $\Gamma_{K^*}$, $M_N$ is the average mass of the nucleons, and $M_{\bar{p}\Lambda}$ is the invariant mass of the $\bar{p}\Lambda$ system.  The parameter $\alpha$ is dimensionless because we have included the $M_N^2$  in the numerator of Eq.~(\ref{eq:amp_kstar}).

The amplitude for the intermediate $N(2300)$ term is
\begin{eqnarray}
\mathcal{M}^{N(2300)}&=&\frac{V_p\beta /M_{N(2300)}}{M_{\Lambda K^+}-M_{N(2300)}+i\Gamma_{N(2300)}/2} \nonumber \\
&& \times<m_{\bar{p}}|\vec{\sigma} \cdot \vec{p}_{\bar{p}}|m_{N(2300)}><m_{N(2300)}|\vec{\sigma}\cdot \vec{p}_\Lambda|m_{\Lambda}>,
\end{eqnarray}
where $\beta$ is the weight of the contribution from the intermediate $N (2300)$ state with a mass of $M_{N(2300)}$ and a width of $\Gamma_{N(2300)}$, $M_{\Lambda K^+}$ is the invariant mass of $\Lambda K^+$ system, $\vec{\sigma}$ is the Pauli matrix, $m_R$ denotes the polarization index of state $R$, and we will sum over the polarizations of the $N(2300)$, $\bar{p}$, and $\Lambda$. $\vec{p}_{\bar{p}}$ and $\vec{p}_\Lambda$ are the three momenta of the $\bar{p}$ and $\Lambda$ in the $\chi_{c0}$ and $K^+\Lambda$ rest frames, respectively.

The amplitude for the intermediate $\Lambda(1520)$ is
\begin{equation}
\mathcal{M}^{\Lambda(1520)}=\frac{V_p\beta'/M^3_{\Lambda(1520)}\times D_{\frac{3}{2}^-}\times D'_{\frac{3}{2}^-}}{M_{\bar{p}K^+}-M_{\Lambda(1520)}+i\Gamma_{\Lambda(1520)}/2}
\end{equation}		
with the term $D_{\frac{3}{2}^-}$ for $\bar{\Lambda}(1520) \to \bar{p}K^+$ vertex
\begin{eqnarray}\label{D}
D_{\frac{3}{2}^-}=\left < m_{\bar{p}}\left |\left[(\tilde{k}_K)_i (\tilde{k}_K)_j-\frac{1}{3}\tilde{k}^2_K\delta_{ij}\right ]\sigma_i\sigma_j\right |m_{\bar{\Lambda}(1520)}\right >,
\end{eqnarray}
and the term $D^{\prime}_{\frac{3}{2}^-}$ for $\chi_{c0} \to\Lambda \bar{\Lambda}(1520)$ vertex
\begin{eqnarray}\label{D}
D^{\prime}_{\frac{3}{2}^-}=
\left <m_{\bar\Lambda(1520)}\left |\left [(\tilde{p}_{\Lambda})_i(\tilde{p}_{\Lambda})_j-\frac{1}{3}\tilde{p}^2_{\Lambda}\delta_{ij}\right ]\sigma_i\sigma_j\right |m_{{\Lambda}}\right >,
\end{eqnarray}
where $\beta'$ corresponds to the weight of the contribution from the intermediate $\Lambda(1520)$ resonance with a mass of $M_{\Lambda(1520)}$ and a width of $\Gamma_{\Lambda(1520)}$, $M_{\bar{p}K^+}$ is the invariant mass of $\bar{p}K^+$ system, and $\tilde{k}_K$ and $\tilde{p}_{\Lambda}$ are the three momenta of $K^+$ and  $\Lambda$  in the $\bar{p}K^+$ and $\chi_{c0}$ rest frames, respectively.

Finally, the invariant mass distributions of $\chi_{c0} \to \bar{p} K^+\Lambda $ read
\begin{eqnarray}\label{dgdm}
\frac{d^2\Gamma}{dM_{\bar{p}K^+}^2 dM_{\bar{p}\Lambda}^2} = \frac{1}{(2\pi)^3} \frac{4M_{\bar{p}}M_{\Lambda}}{32M_{\chi_{c0}}^3}\left|{\cal M}^{\rm total}\right|^2,\label{eq:dw}
\end{eqnarray}
\begin{eqnarray}
\frac{d^2\Gamma}{dM_{\Lambda K^+}^2 dM_{\bar{p}\Lambda}^2} = \frac{1}{(2\pi)^3} \frac{4M_{\bar{p}}M_{\Lambda}}{32M_{\chi_{c0}}^3}\left|{\cal M}^{\rm total}\right|^2,\label{eq:dw1}
\end{eqnarray}
where $M_{\bar{p}}$, $M_{\Lambda}$ , and $M_{\chi_{c0}}$ are the masses of $\bar{p}$, $\Lambda$, and $\chi_{c0}$, respectively. Since there is no interference between the different partial waves, the $|\mathcal{M}^{\rm total}|^2$ in Eqs.(\ref{eq:dw}) and (\ref{eq:dw1}) can be substituted by
\begin{eqnarray}
|\mathcal{M}^{\rm total}|^2&=&V^2_p \left| 1+ \frac{\alpha M^2_N}{M_{\bar{p}\Lambda}^2-M_{K^*}^2+i M_{K^*}{\Gamma_{K^*}}} \right|^2 \nonumber \\
&& +V^2_p \left|\frac{\beta \tilde{p}_{\Lambda} \vec{p}_{\bar p}/M_{N(2300)}}{M_{\Lambda K^+}-M_{N(2300)}+i\Gamma_{N(2300)}/2}\right|^2  +V^2_p \frac{|\tilde{k}_K|^4 |\vec{p}_{\Lambda}|^4}{M_{\Lambda(1520)}^6}  \left|\frac{\beta'}{M_{\bar{p}K^+}-M_{\Lambda(1520)}+i\Gamma_{\Lambda(1520)}/2}\right|^2 .
\end{eqnarray}

The $\bar{p}K^+$ and $\Lambda K^+ $ mass distributions can be obtained by integrating $M_{\bar{p}\Lambda}$ in Eqs.~(\ref{eq:dw}) and (\ref{eq:dw1}) respectively, and the $\bar{p}\Lambda$ mass distributions can be obtained by integrating $M_{\bar{p}K^+}$ in Eq.~(\ref{eq:dw}).
For a given value of  $M_{12}^2$, the range of $M_{23}^2$ is defined as
\begin{eqnarray}\label{daliz}
(M_{23})^2_{\rm max}=(E_{2}^*+E^*_3)^2-\left(\sqrt{{E_{2}^{*2}}-M^2_{2}}-\sqrt{E^{*2}_{3}-M^2_{3}}\right)^2,\nonumber\\
(M_{23})^2_{\rm min}=(E_{2}^*+E^*_3)^2-\left(\sqrt{{E_{2}^{*2}}-M_{2}^2}+\sqrt{E_{3}^{*2}-M_{3}^2}\right)^2,
\end{eqnarray}
where $E_{2}^*$ and $E_{3}^*$ are the energies of particles 2 and 3 in the rest frame of  particles 1 and 2, respectively, and $M_1$ and $M_2$ are the masses of particles 1 and 2, respectively. The masses and widths of the baryons and mesons except for $N(2300)$ involved in this work are taken form PDG~\cite{PDG2018} as follows, $M_{\bar{p}}=938.272$~MeV, $M_{\chi_{c0}}=3414.71$~MeV, $M_{K^+}=493.677$~MeV, $M_\Lambda=1115.683$~MeV, $M_{K(1830)}=1874$~MeV, $\Gamma_{K(1830)}=168$~MeV, $M_{\Lambda(1520)}=1519.5$~MeV, and $\Gamma_{\Lambda(1520)}=15.6$~MeV. For the $N(2300)$, we don't take the measured mass and width due to the larger uncertainties, and take $M_{N(2300)}$  and $\Gamma_{N(2300)}$ as free parameters.


It should be stressed that we do not consider unitarity constraints on the full amplitude $\mathcal{M}^{\rm total}$ of Eq.~(\ref{eq:fullamp}). One has to do further loops with the other two final particles to account for the three-body unitarity. For instance, the $\bar{p}$ and $K^+$ of Fig.~\ref{fig:feyn}(b) can undergo the rescattering. In this case, the invariant mass  of $K(1830)$ gets a distribution and we no longer have a resonant contribution. It implies that those contributions are very small usually, with only one exception that the loop gives rise to a triangle singularity where the $K(1830)$, $K^+$, and $\bar{p}$ are placed on shell in the loop. It is easy to test that there is no triangle singularity following Refs.~\cite{Bayar:2016ftu,Wang:2016dtb,Liang:2019jtr}. In summary, the correction of implementing the tree body unitarity is extremely small and can be neglected in this work.

\section{results and discussions}
\label{sec:results}
\begin{table}
\begin{center}
\caption{ \label{tab:para2}The model parameters obtained by fitting to the BESIII measurements~\cite{Ablikim:2012ff}.}
\begin{tabular}{c  c  c  c  c  c  c c}
\hline\hline
 Parameter   &$\alpha$       &$\beta$        &$\beta^{'}$  & $M_{N(2300)}$  & $\Gamma_{N(2300)}$    & $V_p$                 &$V_{p}^{'}$\\
\hline
  value       & 104.0   &45.6         &6.8    &2354.6     &252.0               &0.010                 &0.022\\
 error       & 2.1  &3.2         &0.5   &18.5      &3.3          &0.001 &0.001\\
\hline\hline
\end{tabular}
\end{center}
\end{table}

With the above formalism, we will fit our model to the $\bar{p}K^+$, $\Lambda K^+$, and $\bar{p}\Lambda$ mass distributions of the events reported by the BESIII Collaboration~~\cite{Ablikim:2012ff}. It should be pointed out that the $\bar{p}K^+$ and $\Lambda K^+$ mass distributions are not corrected by the detector efficiency\footnote{The data of the $\bar{p}K^+$ and $\Lambda K^+$ mass distributions are not corrected by the detector efficiency. We have communicated with Wen-Biao Yan and Cong Geng, the two of authors of Ref.~\cite{Ablikim:2012ff}.  The curves of the detector efficiency distributions for $\bar{p}K^+$ and $\Lambda K^+$  approximate to be flat, and there is no fine structure in the efficiency distributions.}, but the  $\bar{p}\Lambda$ mass distribution is given with the acceptance correction. In order to directly compare our results with the BESIII measurements, we take two different normalization factors in our fit, $V_p$ for the $\bar{p}K^+$/$\Lambda K^+$ mass distribution and $V'_p$ for the $\bar{p}\Lambda$ mass distribution. There are seven model parameters, 1) $\alpha$, the weight of the contribution from the intermediate $K(1830)$ state, 2) $\beta$, the weight of the contribution from the intermediate $N(2300)$ state, 3) $\beta'$, the weight of the intermediate $\Lambda(1520)$ state,  4) the mass and width of the $N(2300)$ state,  5) the unknown
normalization factor $V_p$ for the $\bar{p}K^+$/$\Lambda K^+$ invariant mass distribution, and 6) the unknown
normalization factor $V'_p$ for the $\bar{p}\Lambda$ invariant mass distribution.

\begin{figure}[h]
  \centering
  {\includegraphics[width=0.45\textwidth]{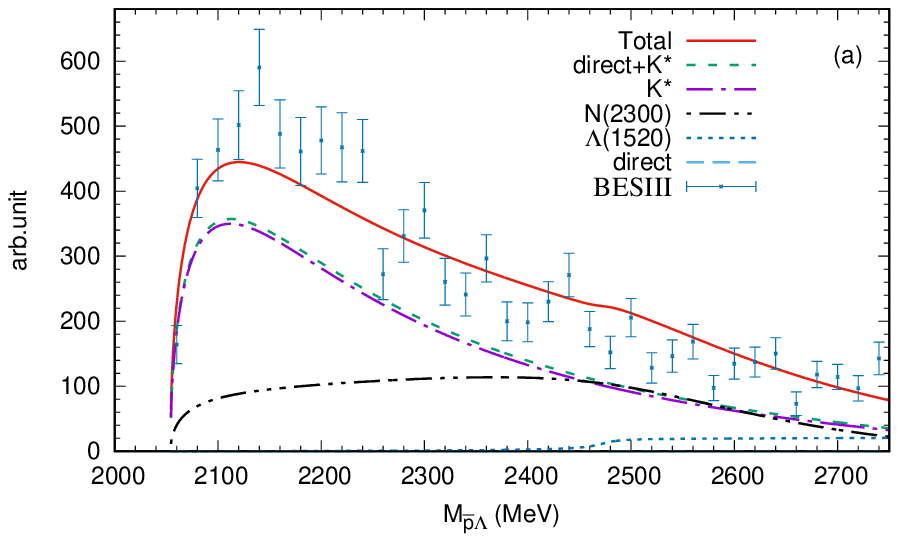}}
  {\includegraphics[width=0.45\textwidth]{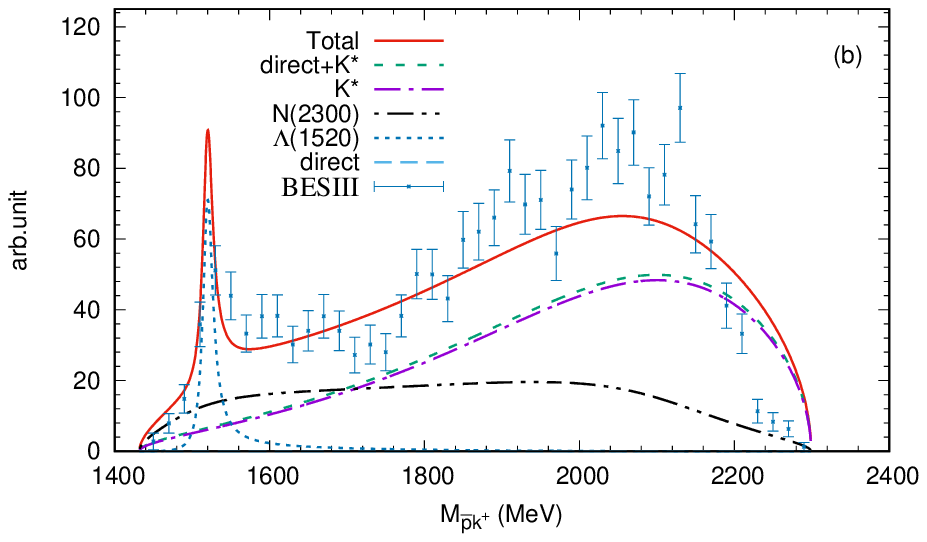}}
  {\includegraphics[width=0.45\textwidth]{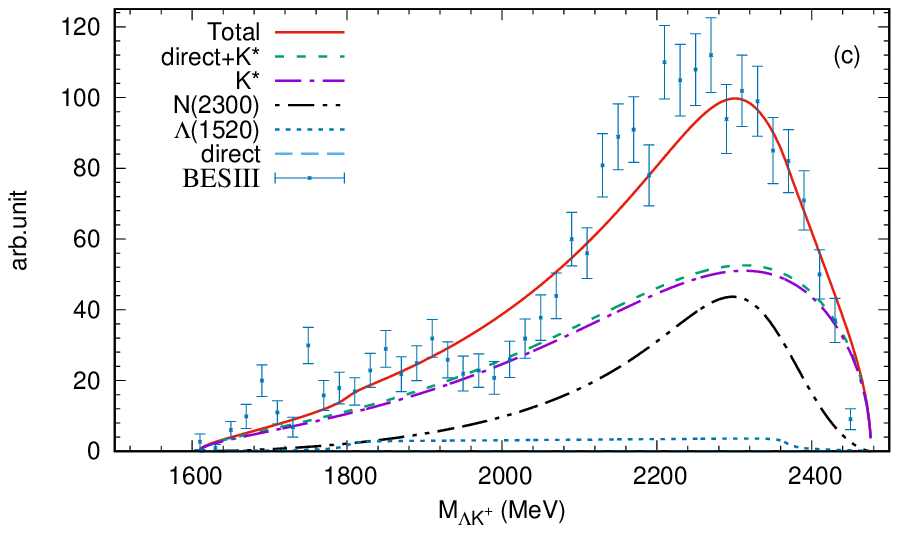}}
   \caption{The $\bar{p}\Lambda$ (a), $\bar{p}K^+$ (b), and $\Lambda K^+$ (c) mass distributions for the reaction $\chi_{c0}\to \bar{p}K^+\Lambda$. The curves labeled as `direct', `$K^*$', `$N(2300)$', and `$\Lambda(1520)$' show the contributions of Figs.~\ref{fig:feyn}(a), (b), (c), and (d), respectively, and the curves labeled as `direct+$K^*$' are the contributions of the direct term and the intermediate $K(1830)$ state in Eq.~(\ref{eq:amp_kstar}). The `total' curves correspond to the total contribution of Eq.~(\ref{eq:fullamp}). The BESIII data are taken from Ref.~\cite{Ablikim:2012ff}.}
   \label{fig:ds_full}
\end{figure}

With the model presented above, we make a fit to the BESIII measurements, including the $\bar{p}\Lambda$, $\bar{p}K^+$, and $\Lambda K^+$ mass distributions~\cite{Ablikim:2012ff}.  The  $\chi^2/d.o.f$ is $393.7/(115-7)=3.6$, and the fitted parameters are tabulated in Table~\ref{tab:para2}~\footnote{In this table and the following tables, the fitted masses and widthes of the resonances are in MeV.}, where both the fitted mass and width of $N(2300)$ are consistent with the PDG values~\cite{PDG2018} within errors. With the fitted values of the parameters, we calculate the $\bar{p}\Lambda$, $\bar{p}K^+$, and $\Lambda K^+$ mass distributions, and compare our results with the BESIII measurements~\cite{Ablikim:2012ff}, as shown in Fig.~\ref{fig:ds_full}. One can see that our results are in good agreement with the BESIII data, especially in the $\bar{p}\Lambda$ mass distribution  the anomalous enhancement near the $\bar{p}\Lambda$ threshold can be well reproduced. The $K(1830)$ plays an important role for the anomalous enhancement. In addition, Fig.~\ref{fig:ds_full}(c) shows that the peak around $2200\sim 2300$~MeV in the $\Lambda K^+$ mass distribution could mainly result from the $N(2300)$.

\begin{figure}[h]
  \centering
  {\includegraphics[width=0.45\textwidth]{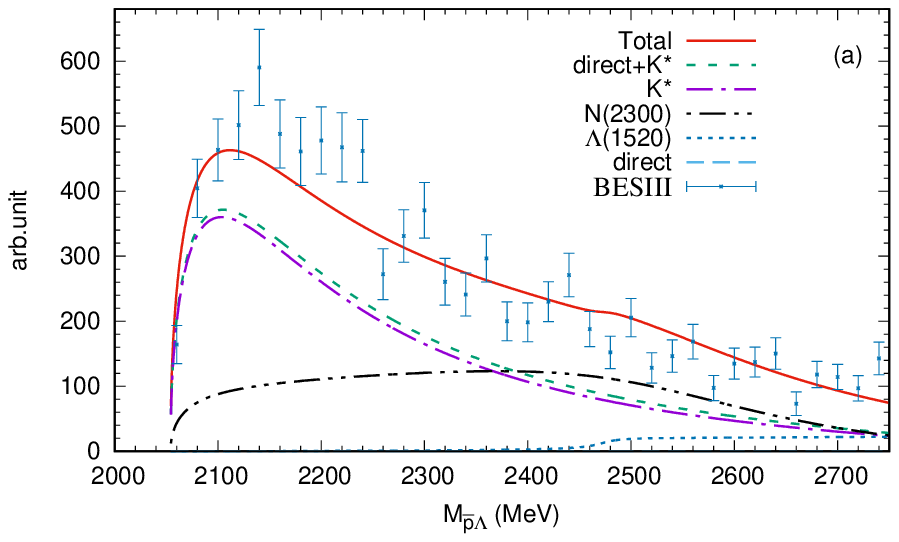}}
  {\includegraphics[width=0.45\textwidth]{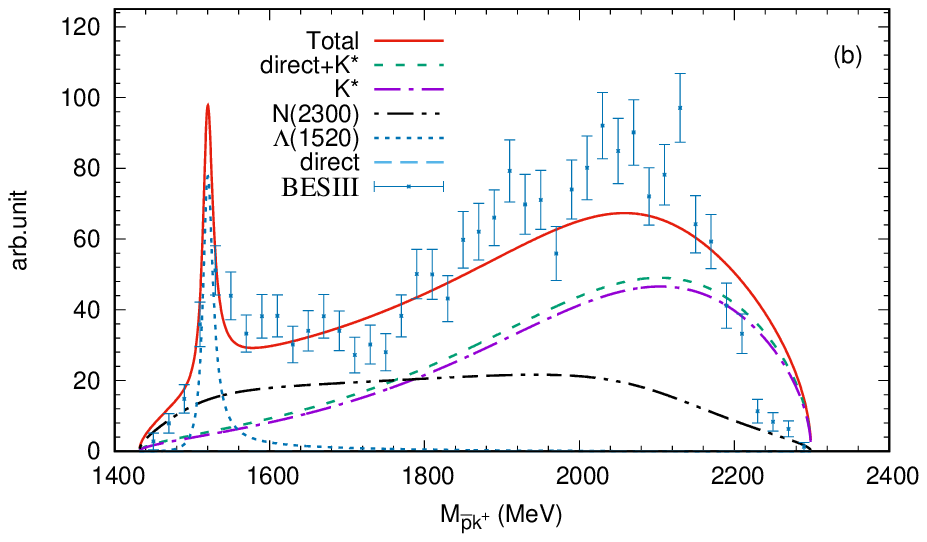}}
  {\includegraphics[width=0.45\textwidth]{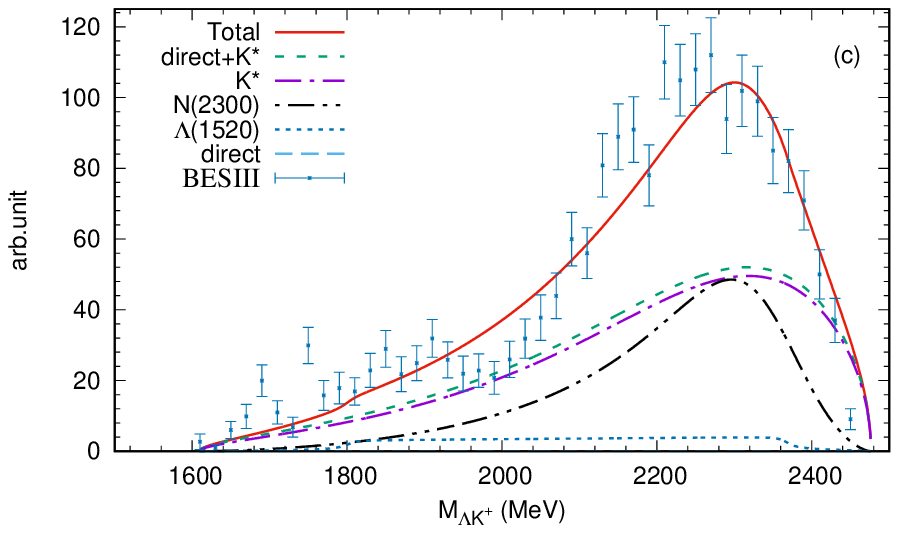}}
   \caption{The $\bar{p}\Lambda$ (a), $\bar{p}K^+$ (b), and $\Lambda K^+$ (c) mass distributions for the reaction $\chi_{c0}\to \bar{p}K^+\Lambda$, by taking the mass  and width of $K(1830)$ as free parameters.
   The explanations of the curves are the same as those of Fig.~\ref{fig:ds_full}.}
   \label{fig:ds_freeK}
\end{figure}

In addition, we also perform the fit again by taking the mass and width of $K(1830)$ as free parameters, since they have large uncertainties ($M_{K(1830)}=1874\pm 43 ^{+59}_{-115}$~MeV, $\Gamma_{K(1830)}=168\pm 90^{+280}_{-104}$~MeV)~\cite{PDG2018}. The $\chi^2/d.o.f$ is $382.5/(115-9)=3.6$, and the fitted parameters are tabulated in Table~\ref{tab:para4}. With the parameters of Table~\ref{tab:para4}, we present the $\bar{p}\Lambda$, $\bar{p}K^+$, and $\Lambda K^+$ mass distributions in Fig.~\ref{fig:ds_freeK}, which are also in good agreement with the BESIII measurements. The $\chi^2/d.o.f$ is the same as that of above fit, and the fitted mass of $K(1830)$ is closer to the $\bar{p}\Lambda$ threshold, which implies that the more precise measurements near the threshold should be useful to constrain the mass  of $K(1830)$.

\begin{table}[h]
\begin{center}
\caption{ \label{tab:para4}The fitted parameters  by taking the mass and width of $K(1830)$ as free parameters.}
\begin{tabular}{c  c  c  c  c  c  c  c  c c}
\hline\hline
 Parameter   &$\alpha$      &$\beta$        &$\beta^{'}$    &$M_{N(2300)}$  &$\Gamma_{N(2300)}$   &$V_p$    &$V_{p}^{'}$  &$M_{K(1830)}$  &$\Gamma_{K(1830)}$\\
\hline
 value       & 47.515       &26.928         &4.0252         &2351.6       &252.01               &0.018         &0.038      &1944.0         &204.53\\
 error       & 14.579       &8.7114         & 1.2638        &16.233       &6.8169               &0.005         &0.011      &13.475         &27.586\\
\hline\hline
\end{tabular}
\end{center}
\end{table}

\begin{figure}[h]
\begin{center}
\includegraphics[width=0.5\textwidth]{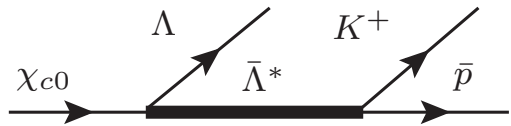}
\caption{The diagram for the contribution from the intermediate high-mass $\Lambda^*$ resonance.}
\label{fig:lambdastar}
\end{center}
\end{figure}

Ref.~\cite{Ablikim:2012ff} has mentioned that the enhancement structure near the $\bar{p}\Lambda$ threshold could simply be explained as an interference effect of high-mass $N^*$ and $\Lambda^*$. In order to check this hypothesis, in addition to the direct diagram contribution, we will only take into account the contributions from the $N(2300)$ and the excited $\Lambda$ resonances $\Lambda^*$'s.  Although there are two states around 2000~MeV, $\Lambda(2100)$ ($7/2^-$) and $\Lambda(2110)$ ($5/2^+$)~\cite{PDG2018}, their contributions are expected to be strongly suppressed since both of them couple to the $\bar{p} K^+$ in $F$-wave. Several $\Lambda^*$ ($1/2^-$) with masses around 2100 MeV have been predicted by the quark model~\cite{Capstick:1986bm}.
For simplicity, we take into account the contribution from one $\Lambda^*$ ($1/2^-$) which couples to the $\bar{p}K^+$ in $S$-wave\footnote{Indeed, if several $\Lambda^*$s with $J^P=1/2^-$ around 2100~MeV are taken into account, the contributions can be described with the Breit-Winger form of one $\Lambda^*$, by adjusting its mass and width. }, as shown in Fig.~\ref{fig:lambdastar}. The corresponding amplitude can be expressed as,
\begin{equation}
\mathcal{M}^{\Lambda^*}=\frac{V_p\alpha' M_{N}}{M_{\bar{p}K^+}-M_{\Lambda^*}+i\Gamma_{\Lambda^*}/2},
\end{equation}
where $\alpha'$ is the weight of the contribution from the intermediate $\Lambda^*$ resonance with a mass of $M_{\Lambda^*}$ and a width of $\Gamma_{\Lambda^*}$. The full amplitude can be rewritten as,
\begin{equation}
(\mathcal{M}')^{\rm total}=\mathcal{M}^{\rm direct}+\mathcal{M}^{N(2300)}+\mathcal{M}^{\Lambda^*}.\label{eq:fullamp2}
\end{equation}

\begin{figure}[h]
  \centering
 {\includegraphics[width=0.45\textwidth]{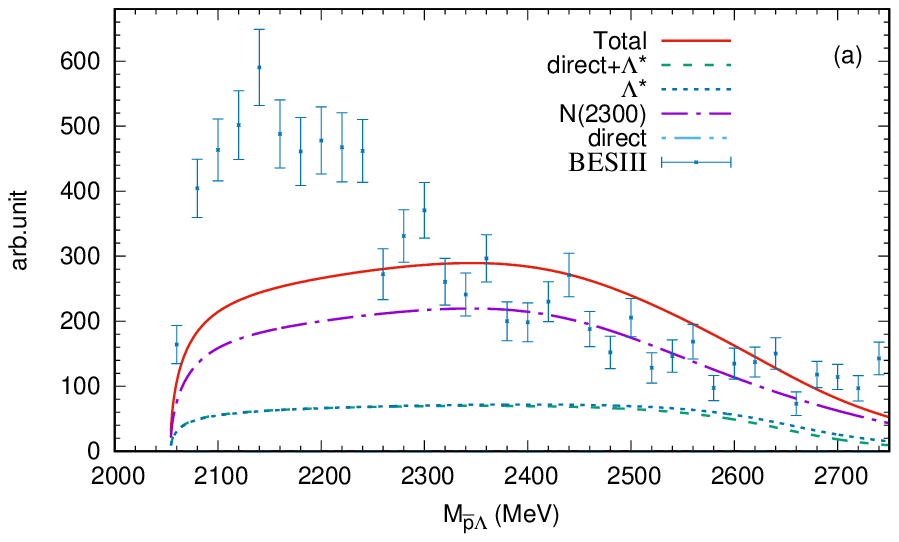}}
 {\includegraphics[width=0.45\textwidth]{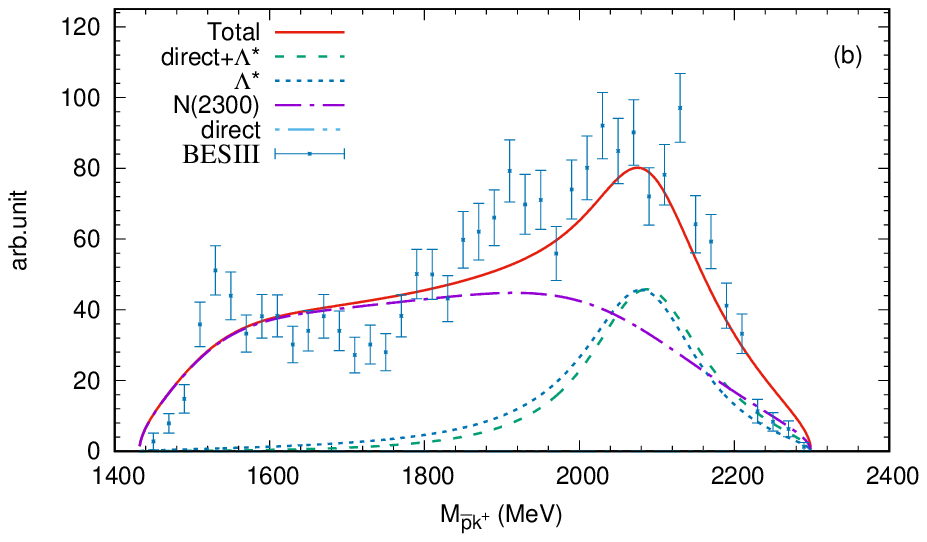}}
  {\includegraphics[width=0.45\textwidth]{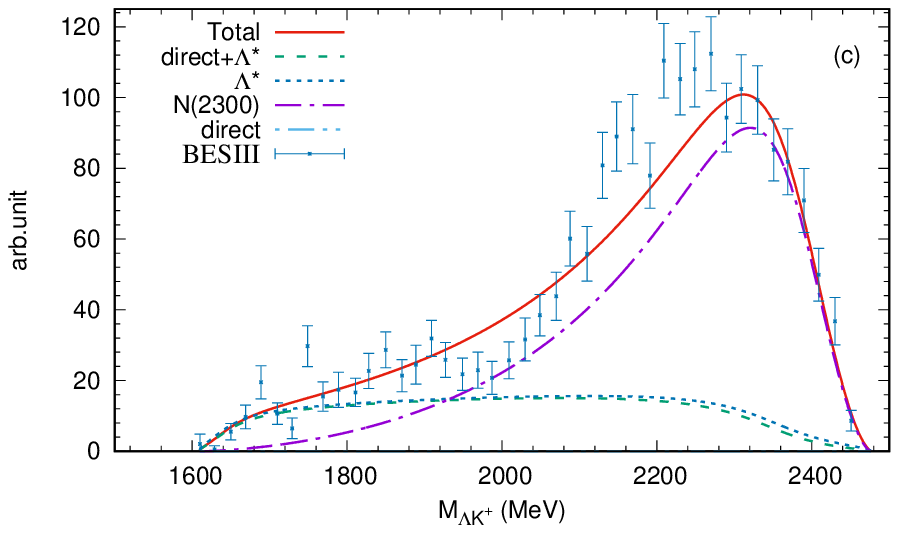}}
   \caption{The $\bar{p}\Lambda$ (a), $\bar{p}K^+$ (b), and $\Lambda K^+$ (c) mass distributions including the $N(2300)$ and high mass $\Lambda$ state.  The curves labeled as `direct',  `$N(2300)$', and `$\Lambda^*$' show the contributions of the Figs.~\ref{fig:feyn}(a),  (c), and Fig.~\ref{fig:lambdastar}, respectively, and the curves labeled as `direct+$\Lambda^*$' are the contribution of the direct term and the intermediate $\Lambda^*$ state,. The `total' curves correspond to the total contribution of Eq.~(\ref{eq:fullamp2}). The BESIII data are taken from Ref.~\cite{Ablikim:2012ff}. }
   \label{fig:ds_full2}
\end{figure}

\begin{figure}[h]
  \centering
  {\includegraphics[width=0.45\textwidth]{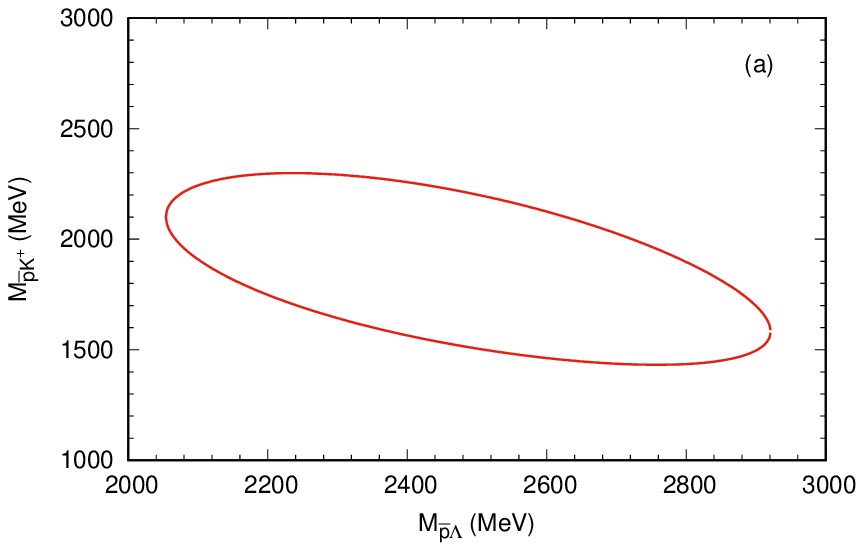}}
  {\includegraphics[width=0.45\textwidth]{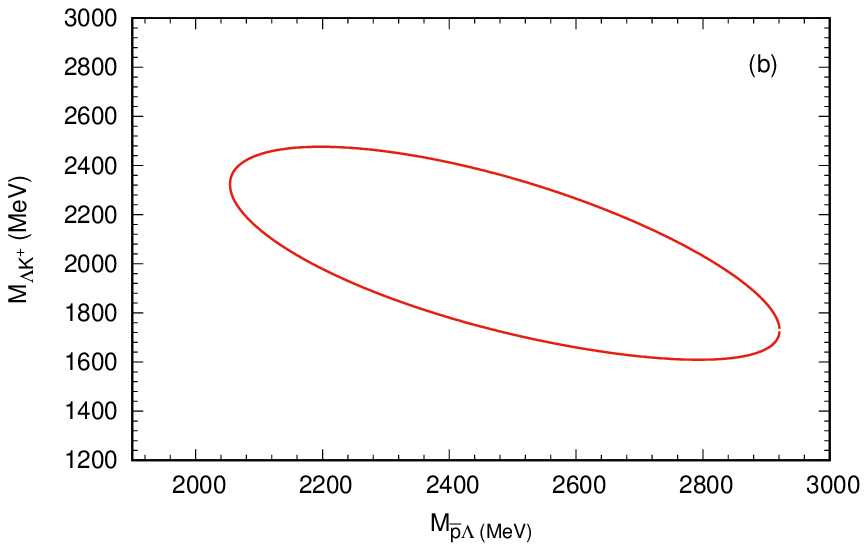}}
   \caption{{The Daliz plots for the $\chi_{c0} \rightarrow \bar{p} K^+ \Lambda$ }}
   \label{fig:daliz}
\end{figure}

In this case, we have eight parameters, 1) $\alpha'$, the weight of the contribution from the intermediate $\Lambda^*$ resonance, 2) $\beta$, the weight of the contribution from the intermediate $N(2300)$, 3) the mass and the width of the $\Lambda^*$, 4) the mass and width of the $N(2300)$, 5) two unknown normalization factors $V_p$ and $V^{\prime}_p$. With the amplitude of Eq.~(\ref{eq:fullamp2}),  we make a fit to the BESIII data~\cite{Ablikim:2012ff}, and find $\chi^2/d.o.f=5.54$, which is larger than that of the above cases. The fitted parameters are tabulated in Table~\ref{tab:para3}.
We also present the  $\bar{p}\Lambda$, $\bar{p}K^+$, and $\Lambda K^+$ mass distributions in Fig.~\ref{fig:ds_full2}. Although the $\bar{p}K^+$ and $\Lambda K^+$ mass distributions can be well reproduced, the anomalous enhancement near the $\bar{p}\Lambda$ threshold is not found in the $\bar{p}\Lambda$ mass distribution. This can be explained by the Dalitz plots of the $\chi_{c0}\to \bar{p}\Lambda K^+$ as shown in Fig.~\ref{fig:daliz}. It shows that the high mass $N^*$ and $\Lambda^*$ give the contributions in the energy regions of $2100\sim 2600$~MeV of the $\bar{p}\Lambda$ mass distribution, not only in the energy regions near the $\bar{p}\Lambda$ threshold, which is in agreement with Fig.~\ref{fig:ds_full2}(a) (see the curves labeled as `$N(2300)$' and `$\Lambda^*$').
 Based on the partial wave analysis, the BESIII Collaboration has also concluded that the enhancement in the $J/\psi\to \bar{p}\Lambda K^+ + c.c.$ cannot be due to the interference effects between high-mass $N^*$'s and $\Lambda^*$'s~\cite{Ablikim:2004dj}

\begin{table}[h]
\begin{center}
\caption{ \label{tab:para3}The model parameters obtained by fitting to the BESIII measurements~\cite{Ablikim:2012ff}, taking into account the interference of $N(2300)$ and $\Lambda^*(1/2)$.}
\begin{tabular}{c  c  c  c  c  c  c  c  c}
\hline\hline
 Parameter   &$\alpha^{'}$    &$\beta$     &$V_p$     &$V_{p}^{'}$    &$M_{\Lambda^*}$   &$\Gamma_{\Lambda^*}$  &$M_{N(2300)}$   &$\Gamma_{N(2300)}$\\
\hline
 value       &1.16    &17.1     &0.046     &0.093          &2085.1            &183.2                &2402.1          &252.0\\
 error       &0.13    &1.4     &0.004     &0.008          &8.9            &20.1              &12.2 &2.3\\
\hline\hline
\end{tabular}
\end{center}
\end{table}

\begin{figure}[h]
  \centering
  \subfigure[]{\includegraphics[width=0.45\textwidth]{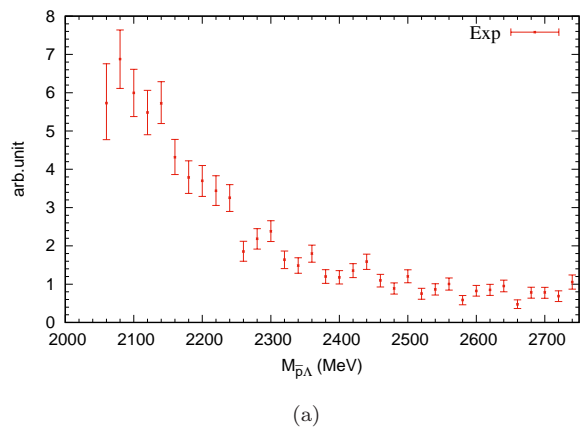}}

   \caption{Data for $d\Gamma/dM_{\bar{p}\Lambda}$ divided by the phase space of Eq.~(\ref{eq:dw}).}
   \label{fig:removephase}
\end{figure}

Finally, in Fig.~\ref{fig:removephase}, we show that the results for $|\mathcal{M}|^2$ extracted from the BESIII data~\cite{Ablikim:2012ff}, dividing the measured $\bar{p}\Lambda$ mass distribution by the phase space factor of Eq.~(\ref{eq:dw}). One can see that the peak should be below the $\bar{p}\Lambda$ threshold if the first point of Fig.~\ref{fig:removephase} is neglected because of the limited statistics.
Thus, if the anomalous enhancements near the $\bar{p}\Lambda$ threshold   in the processes $\chi_{c0}\to \bar{p}K^+\Lambda $~\cite{Ablikim:2012ff}, $J/\psi \to p K^-\bar{\Lambda}+c.c.$, $\psi' \to p K^-\bar{\Lambda}+c.c.$~\cite{Ablikim:2004dj}, and $B^0\to p \bar{\Lambda} \pi^-$~\cite{Wang:2003yi} are due to the $K^*$ resonance, there will be a peak structure around 1900~MeV in the $K\pi$, $K\pi\pi$, $K\phi$ modes of the processes $\chi_{c0}, J/\psi, \psi'\to K K\pi, KK\pi\pi, KK\phi$. Searching for the structure in those processes would be helpful to understand the anomalous enhancements near the $\bar{p}\Lambda$ threshold.

\section{Summary}
\label{sec:summary}
In this work, we have analyzed the anomalous enhancement near the $\bar{p}\Lambda$ threshold in the $\chi_{c0}\to \bar{p}K^+\Lambda $ reaction measured by the BESIII collaboration~\cite{Ablikim:2012ff}. Our results for the $\bar{p}\Lambda$, $\bar{p}K^+$, and $\Lambda K^+$ mass distributions are in good agreement with the BESIII measurements~\cite{Ablikim:2012ff}. We find that the anomalous enhancement near the $\bar{p}\Lambda$ threshold is mainly due to the contribution of the $K(1830)$. We have also shown that interference of the high-mass $N^*$ and $\Lambda^*$ can not reproduce the anomalous enhancement near the $\bar{p}\Lambda$ threshold.

It is an usual way to identify the peak structure in the mass distribution as a resonance, such as $N(2300)$ and $\Lambda(1520)$ in this work, although sometimes the kinematic effects can also give rise to the peak or cusp structure~\cite{Guo:2017jvc,Guo:2014iya,Dai:2018nmw,Wang:2018djr,Wang:2017mrt}. However, one should be very careful for the enhancement near the threshold, which usually indicates the existence of the bound state or resonance below the threshold. One purpose of the work is to check whether the BESIII mearsurements can be described or not by introducing the resonance $K(1830)$ below the $\bar{p}\Lambda$ threshold.

We suggest to confirm the peak structure around 1900~MeV in the $K\pi$, $K\pi\pi$, and $K\phi$ modes of the processes $\chi_{c0}, J/\psi, \psi'\to K K\pi, KK\pi\pi, KK\phi$, which will help to understand the anomalous enhancement near the $\bar{p}\Lambda$ threshold.

\section*{Acknowledgements}
We would like to acknowledge the fruitful discussions with Eulogio Oset, Ju-Jun Xie, and Li-Sheng Geng.
This work is partly supported by the National Natural Science Foundation of China under Grant No. 11505158, the Key Research Projects of Henan Higher Education Institutions (No. 20A140027),  the Fundamental Research Cultivation Fund for Young Teachers of Zhengzhou University, and the Academic Improvement Project of Zhengzhou University.

\end{document}